\documentclass[prd,twocolumn]{revtex4}
\pdfoutput=1
\usepackage{amssymb,latexsym}
\usepackage{amsmath,amsbsy,bbm}
\usepackage{ifpdf}
\usepackage{epsfig,bm}
\usepackage{graphicx,comment}
\usepackage{color}
\usepackage{soul}
\usepackage{mathtools}
\usepackage{comment}
\usepackage[normalem]{ulem}
\begin{document}

\title{Machine learning phases of an Abelian gauge theory}
\author{Jhao-Hong Peng}
\affiliation{Department of Physics, Box 90305, Duke University, Durham, NC 27708, USA}
\affiliation{Department of Physics, National Taiwan Normal University,
88, Sec.4, Ting-Chou Rd., Taipei 116, Taiwan}
\author{Yuan-Heng Tseng}
\affiliation{Department of Physics, National Taiwan Normal University,
88, Sec.4, Ting-Chou Rd., Taipei 116, Taiwan}
\author{Fu-Jiun Jiang}
\email[]{fjjiang@ntnu.edu.tw}
\affiliation{Department of Physics, National Taiwan Normal University,
88, Sec.4, Ting-Chou Rd., Taipei 116, Taiwan}

\begin{abstract}
  The phase transition of the two-dimensional $U(1)$ quantum link model on the triangular lattice
  is investigated by employing a supervised neural network (NN) consisting of only one input layer, one hidden layer
  of two neurons, and one output layer. No information on the studied model is used when the
  NN training is conducted. Instead, two artificially made configurations are considered as the training set. Interestingly,
  the obtained NN not only estimates the critical point accurately but
  also uncovers the physics correctly. The results presented here imply
  that a supervised NN, which has a very simple architecture and is trained without
  any input
  from the investigated model, can identify the targeted phase structure with
  high precision.

\end{abstract}

\maketitle

\section{Introduction}\vskip-0.3cm

Machine learning (ML) has attracted much attention in the physics community
recently. Many ML techniques have been applied to study physics systems
and these applications have reached various degrees of success \cite{Rup12,Sny12,Baldi:2014pta,Mnih:2015jgp,Searcy:2015apa,Oht16,Att16,Hoyle:2015yha,Mott:2017xdb,Car16,Tan16,Nie16,Den17,Zha17,Hu17,Tubiana:2016zpw,Kol17,Li18,Chn18,Lu18,But18,Pang:2016vdc,Sha18,Rod19,Cavaglia:2018xjq,Zha19,Don19,Conangla:2018nnn,Tan20.1,Lidiak:2020vgk,Tan20.2,Shalloo:2020nhu,Geilhufe:2020nbl,Larkoski:2017jix,Aad:2020cws,Morgan:2020wvf,Nicoli:2020njz,Tan21,Hse22}.
For instance, the methods of neural networks (NN) are demonstrated to be able to
identify phases of matter with high accuracy \cite{Oht16,Car16,Tan16,Nie16,Den17,Zha17,Hu17,Li18,Chn18,Rod19,Zha19,Don19,Tan20.1,Lidiak:2020vgk,Tan20.2,Tan21}.

Conventionally, one needs to train a NN with certain information about the targeted system
in order to carry out a NN study of the associated phase transition. Apart from
using relevant physical quantities for the training, the NN considered
for such kind of investigation typically has a very complicated architecture.
As a result, it takes a huge amount of computing resources
to adopt the NN approach to study a phase transition. Because of this,
the applications of NN are limited, in particular, to systems of
small sizes.

Under such a situation, it is of little motivation to use NN instead of the
traditional methods when phase transitions are concerned. Unless certain
drawbacks of NN approaches are overcome, NN is hindered from
being widely adopted in reality.

In Ref.~\cite{Hse22}, an autoencoder (AE) and a generative adversarial network (GAN)
have been constructed. In particular, each of these two unsupervised NN
consists of only one input layer, one hidden layer of 2 neurons, and
one output layer. Moreover, these two unsupervised NNs are trained on a
one-dimensional (1D) lattice of 200 sites and the training set is composed of
two artificially made configurations. It has been shown that these
two unsupervised NN successfully estimate the critical points of
the three-dimensional (3D) classical $O(3)$ model, the
two-dimensional (2D) generalized $XY$ model, the 2D
ferromagnetic Potts model, and the 1D Bose-Hubbard
model with high precision. A benchmark calculation indicates that
a factor of few thousand in the speed of calculation
is gained for the unconventional
AE and GANs when they are compared with the performance of
the standard NN. A similar idea has been applied to supervised NN
as well \cite{Tan21}.

In this study, we build a supervised NN made up of one input layer,
one hidden layer of two neurons, and one output layer.
In addition, we train the supervised NN on a 1D lattice of 200 sites
and use two artificially made configurations as the training set.
The training process takes only 24 seconds.
The constructed NN is then employed to study the phase structure
of the 2D $U(1)$ quantum link model, which is an Abelian gauge theory,
on the triangular lattice.

The quantum link models are generalizations of the
Wilson lattice gauge theory \cite{Wil74,Hor81,Orl90,Cha98,Bro99}.
In addition to serving as a regularization of quantum chromodynamics (QCD),
these models have very rich physical phenomena such as the splitting of
confining strings into individual strands carrying fractionalized electric
flux \cite{Ban13,Ban18,Ban22}.

The spontaneous symmetry breaking patterns of the
2D $U(1)$ quantum link model on the triangular lattice before and beyond the
phase transition are nontrivial. In other words, the model has an exotic type
of phase transition \cite{Ban22}. As a result, it will be intriguing to examine whether the employed
NN, which is trained without any input from the targeted model,
can identify the physics of the considered system correctly.

Interestingly, although the employed supervised NN has only one hidden
layer of 2 neurons and is trained without any
information of the considered model, it has determined the
associated critical point accurately. Moreover, for both phases,
the used NN correctly predicts
whether the sublattices are ordered or disordered. This is remarkable
because the NN does not know anything about the studied
model. The outcomes shown here as well as that in Refs.~\cite{Tan21,Hse22} strongly
suggest that training with the input
of the considered system is not necessary for a NN to be able to
distinguish various phases. In particular, NNs trained with two artificial
configurations, such as the one presented here and those in Refs.~\cite{Tan21,Hse22},
are universal since they can be applied to determine the critical points of
many models successfully.

The rest of the paper is organized as follows. After the introduction,
the 2D $U(1)$ quantum link model and the employed NN are introduced briefly
in Secs.~II and III, respectively. The NN outcomes are presented in Sec.~IV.
We conclude our investigation in Sec.~V.

\section{The considered model}

\begin{figure}
  \vskip-0.5cm

       \includegraphics[width=0.3\textwidth]{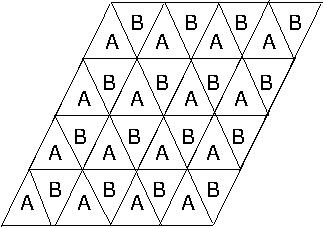}        
        \vskip-0.2cm
        \caption{The rhombic lattice considered in our study. The rhombic
          lattice consists of two sublattices $A$ and $B$.}
        \label{fig0}
\end{figure}

The Hamiltonian of the $U(1)$ quantum link model on the triangular
lattice takes the form
\begin{equation}
H\!=\!\sum_\bigtriangleup H_\bigtriangleup\!=\!
- J \sum_\bigtriangleup \left[U_\bigtriangleup + U_\bigtriangleup^\dagger - 
\lambda (U_\bigtriangleup + U_\bigtriangleup^\dagger)^2\right],
\label{ham}
\end{equation}
where $U_\bigtriangleup = U_{x_1x_2} U_{x_2x_3} U_{x_3x_1}$ is an operator associated
with the parallel transport around a triangular plaquette $\bigtriangleup$.
Here the quantum link operator $U_{x_1x_2}$ connecting nearest-neighbor
sites $x_1$ and $x_2$ is given by
$U_{x_1x_2} = S_{x_1x_2}^1 + i S_{x_1x_2}^2 = S_{x_1x_2}^+$ with $\vec{S}$ being
a quantum spin. A rhombic lattice of side length $L$ with
periodic boundary conditions is considered in our study, see fig.~\ref{fig0}.
Notice
the rhombic lattice consists of two sublattices $A$ and $B$. In addition
to the $U(1)$ gauge symmetry, the system has several global symmetries such
as lattice translations and reflection on a
lattice axis as well.

Instead of working directly with the variable of links,
we consider the dual degrees of freedom, namely the quantum height
variables.
The height variables live on the center of triangular plaquette and
can take values of $+1$ or $-1$.

The construction of a height configuration and its relation to
the original flux configuration is detailed in Ref.~\cite{Ban22}.
Since we are interested in uncovering the relevant physics solely from
the NN data without consulting any information about the original system,
here we will not go into the details of the studied model and will only
briefly summarize the physics and the associated phase structure
of the system.

By setting $J=1$ and varying $\lambda$ in Eq.~\ref{ham}, there exits a
critical value $\lambda_c$ where a phase transition occurs.
The $\lambda_c$ is estimated to be $\sim -0.2150$ in Ref.~\cite{Ban22}.
For $\lambda > \lambda_c$, one of the two sublattices is ordered, implying
the 60 degrees rotation $O$ is broken. When $\lambda < \lambda_c$,
the charge conjugation $C$ is additionally broken, leading to the case
that both sublattices are ordered. The translation symmetry remains unbroken
in both the regions of $\lambda > \lambda_c$ and $\lambda< \lambda_c$. 
Consequently, one has two different nematic phases. The mentioned
phase transition is an exotic weak first-order transition as demonstrated
in Ref.~\cite{Ban22}.

\vskip0.5cm

\begin{figure*}
  \vskip-0.5cm

       \includegraphics[width=0.8\textwidth]{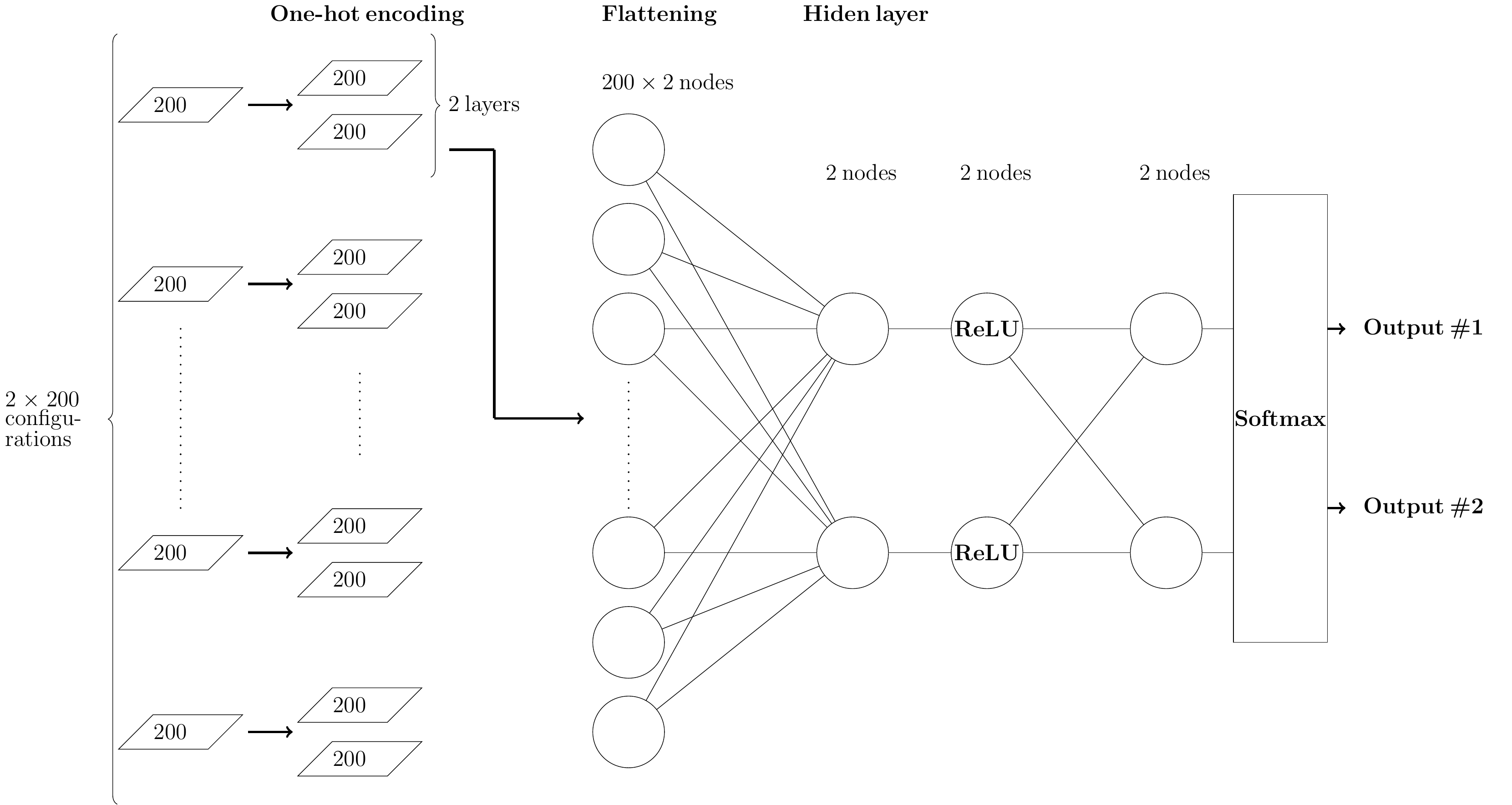}        
        \vskip-0.2cm
        \caption{The supervised neural network considered in this study.}
        \label{fig0.5}
\end{figure*}

\section{The employed NN}

The 1D (supervised) NN employed here is built with Keras and
Tensorflow \cite{kera,tens}. In particular, it is a multilayer
perceptron (MLP) which consists of only one input layer, one hidden
layer of 2 neurons, and one output layer.

Since we want to construct a supervised NN without using any information
from the considered model, the training set employed here consists of
200 copies of two artificially made one-dimensional (1D) lattices of 200 sites.
Particularly, all the sites of one configuration take 1 as their values,
and each element of the other (configuration) is 0. Because the training
set contains only two types of configurations, it is natural to
use $(1,0)$ and $(0,1)$ as the output labels.

The algorithm and optimizer employed in training the NN are minibatch and
adam (the associated learning rate is set to 0.05), respectively. Moreover, activation functions ReLU and softmax
are used in the hidden and output layers, respectively. $L_2$ regularization
is used to avoid overfitting.
The steps of one-hot encoding and flattening are considered as well.
The loss function utilized is categorical crossentropy.
Finally, 800 epochs are conducted and the used batch size is 40. 

A short explanation of the NN terminologies 
mentioned here as well as the details of the training and testing processes
are available in Ref.~\cite{Tan20.2,Tan21}. A cartoon representation
of the employed NN is depicted in fig.~\ref{fig0.5}.

\section{The NN Results}

We have carried out large-scale quantum Monte Carlo simulations and
prepared few to a few thousand height configurations with various system sizes
$L$ and different values of $\lambda$. The simulations are done using
the algorithm introduced in Ref.~\cite{Ban22}. These height configurations
will be used for the testing (prediction) stage. In particular,
for each height configuration, 200 height variables are chosen randomly to build
a configuration that is then considered for the NN prediction.

\subsection{The determination of $\lambda_c$}

\begin{figure}
	\vskip-0.5cm
       \includegraphics[width=0.4\textwidth]{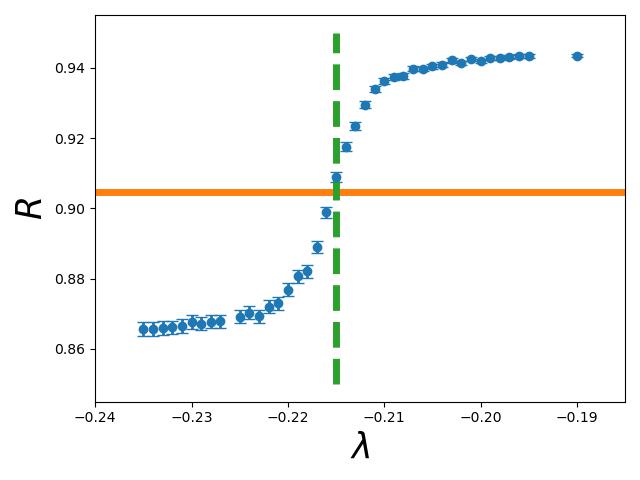}        
        \vskip-0.2cm
        \caption{$R$ as a function of $\lambda$ for $L=24$. The vertical
        	dashed line represents the true $\lambda_c$. The $\lambda$
        	associated with the intersection of the horizontal line and the data
        	of $R$ is the estimated value of $\lambda_c$.}
        \label{fig1}
\end{figure}

\begin{figure}
  \vskip-0.5cm

       \includegraphics[width=0.4\textwidth]{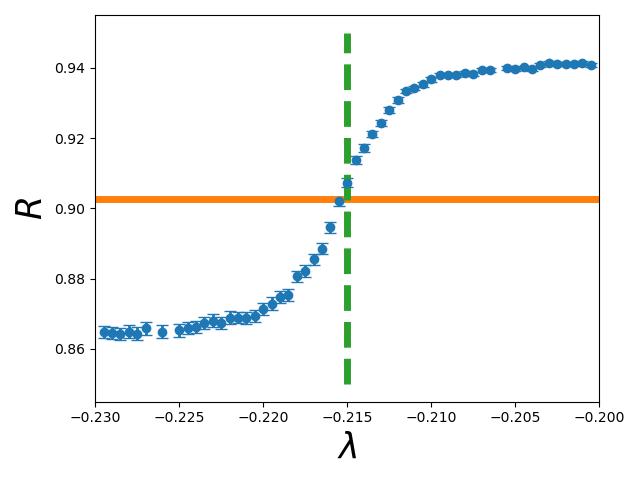}        
        \vskip-0.2cm
        \caption{$R$ as a function of $\lambda$ for $L=32$. The vertical
          dashed line represents the true $\lambda_c$. The $\lambda$
          associated with the intersection of the horizontal line and the data
          of $R$ is the estimated value of $\lambda_c$.}
        \label{fig2}
\end{figure}

\begin{figure*}
  \vskip-0.5cm
  
  \begin{hbox}
      {~~~~~~~~~~~
       \includegraphics[width=0.4\textwidth]{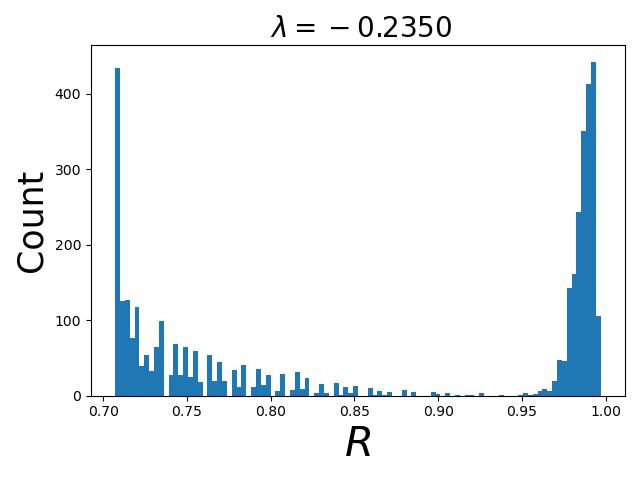}~~~~~~~
       \includegraphics[width=0.4\textwidth]{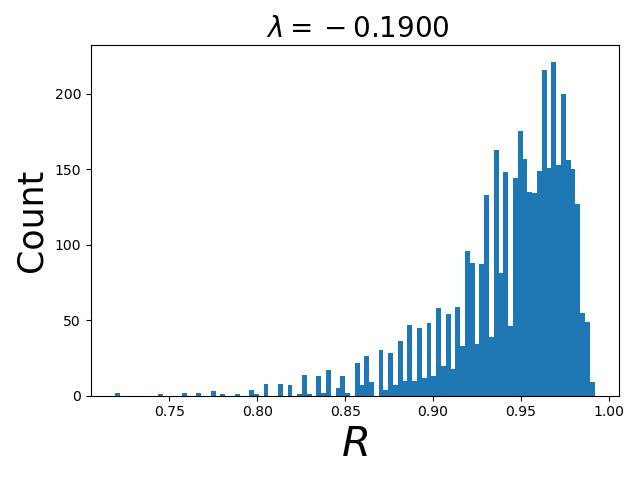} }
  \end{hbox}         
        \vskip-0.2cm
        \caption{The histograms of $R$ for $\lambda = -0.235$ (left) and $\lambda = -0.19$ (right).}
        \label{fig3}
\end{figure*}

The magnitude $R$ of the NN outputs is investigated.
First of all, we would like to examine whether the quantity $R$ can be employed to estimate the critical point.

$R$ as a function of $\lambda$
for $L=24$ is shown in fig.~\ref{fig1}.
By investigating $R$ over various values of $\lambda$, we find that close to
$\lambda_c$ there
is a narrow region $\Lambda$ in $\lambda$ where the magnitude of $R$ has a dramatic
jump. Such a region can be interpreted as the critical region. In addition,
the $R$ associated with those $\lambda$ away from the narrow
region saturate to certain constants. Let the saturation values for $\lambda > \Lambda $ and $\lambda < \Lambda$
be denoted by $R_R$ and $R_L$ respectively. Moreover,
one defines $R_M$ to be $\left(R_R+R_L \right)/2$ ($R_M$ is the horizontal solid
line in fig.~\ref{fig1}). Then the $\lambda$ corresponding to the intersection
of
the horizontal line and the curve of $R$ in fig.~\ref{fig1} matches very well
with the critical point $\lambda_c \sim -0.2150$ (the vertical dashed line in
fig.~\ref{fig1}). This result indicates that such a simple procedure of
computing can lead to a
quite precise estimation of $\lambda_c$. This strategy also works for
the data of $L=32$, see fig.~\ref{fig2} and the related caption. To conclude,
for sufficiently large system sizes, the procedure of computing the intersection
introduced above can be employed to determine the critical point with
reasonable high precision.

\subsection{The histograms of $R$}

After presenting a method of calculating $\lambda_c$, we want to investigate
what information the data of $R$ can reveal. It should be pointed out
that in theory the largest and smallest possible values of $R$ are given by $1$
(When all the height variables take the same value) and
$1/\sqrt{2}$ (When the values of height variables are random), respectively.  

The histogram of $R$ for $\lambda = -0.235 < \lambda_c$ and $L=24$  
is shown in the left panel of fig.~\ref{fig3}, which has a two peaks structure. In particular,
the peaks are located at $1$ and $1/\sqrt{2}$. 
Let us consider two types of configurations.
The first one is that all the chosen height variables have the same value.
The second one is that half of the picked height variables have
the value of 1 (-1) and the other half take -1 (1) as their values.
The former and the later type of configurations will lead to $R = 1$ and $R=1/\sqrt{2}$, respectively. 
To obtained the result shown in the left panel of fig.~\ref{fig3}, these two types of configurations
should occur with equal probability.
In terms of physics, the most probable situation that would
result in such an outcome is that the whole height variables can be put into
two categories (i.e. the underlying lattice is divided into two sublattices)
and each of these two categories is ordered. In other words,
certain symmetry is broken spontaneously in both sublattices.

The histogram of $R$ for $\lambda=-0.19 > \lambda_c$ and $L=24$ is demonstrated in the right panel of
fig.~\ref{fig3}.
The data of $R$ accumulate at the right-hand part of fig.~\ref{fig3}.
This implies that the number of height variables having a particular value
is larger than the number of height variables taking the other value.
If one adopts
the view of dividing the underlying lattices into two sublattices, then this result
can be understood as one sublattice being ordered and the other being disordered.
In particular, certain symmetry is broken spontaneously in one sublattice.

Based on the outcomes shown in figs.~\ref{fig1},~\ref{fig2},~\ref{fig3}, one
concludes that the investigated system has two sublattices $A$ and $B$. This
is in agreement with fig.~\ref{fig0} which indeed has two sublattices.
In other words, one can obtain the correct features of the studied model
solely from the $R$ of the NN outputs. In particular, the NN is trained
without any input from the system.

\subsection{The running histories of $R$}

Having estimated the location of $\lambda_c$ and understood
certain symmetry-breaking patterns of the system, we now turn to explore
the running histories of $R$ on sublattices $A$ and $B$. 

\begin{figure*}
  \vskip-0.5cm
  
  \begin{hbox}
      {
       \includegraphics[width=0.3\textwidth]{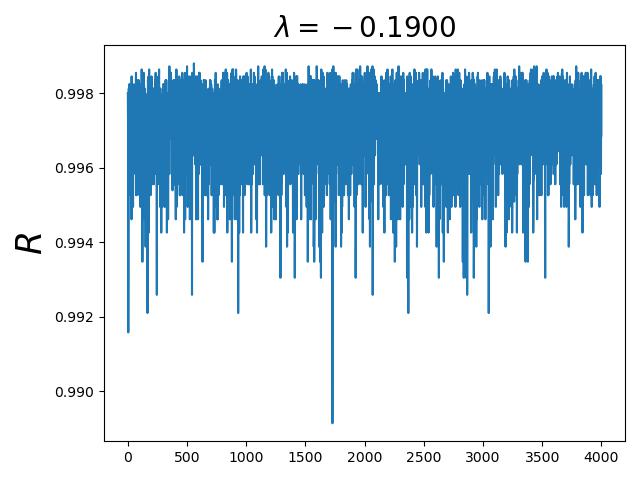}~~
       \includegraphics[width=0.3\textwidth]{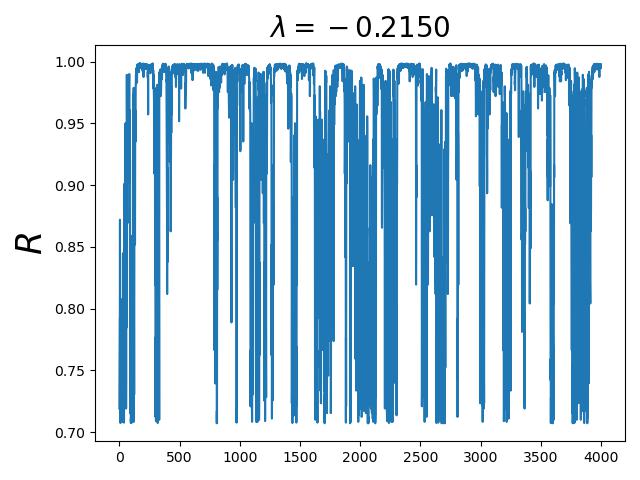}~~
      \includegraphics[width=0.3\textwidth]{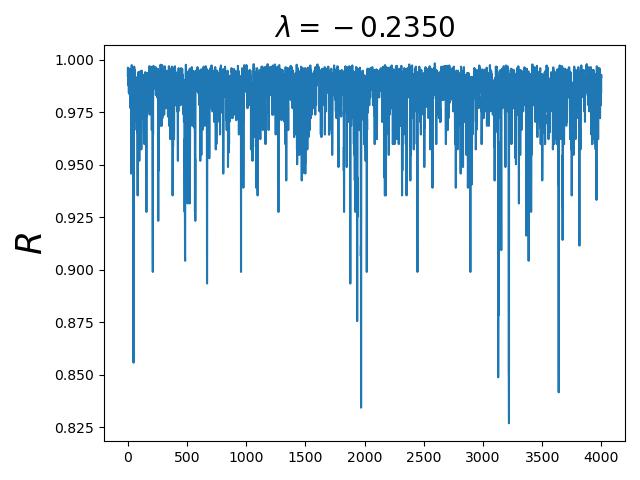}}
  \end{hbox}         
        \vskip-0.2cm
        \caption{The running histories of $R$ of sublattice A for $\lambda = -0.19$ (left), $\lambda = -0.215$ (middle), and $\lambda = -0.235$ (right).}
        \label{fig4}
\end{figure*}

\begin{figure*}
  \vskip-0.5cm
  
  \begin{hbox}
      {
       \includegraphics[width=0.3\textwidth]{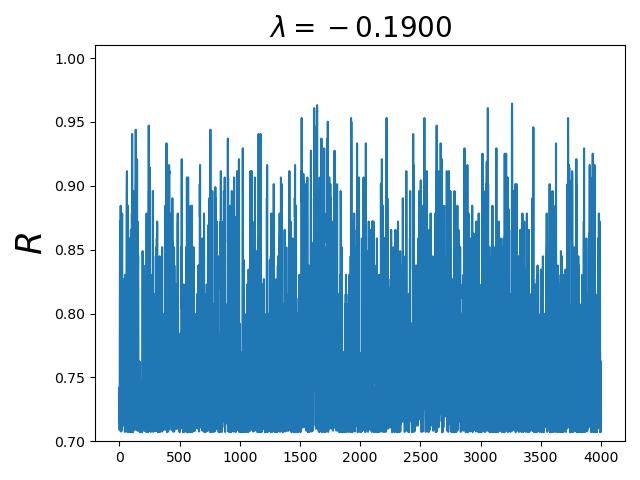}~~
       \includegraphics[width=0.3\textwidth]{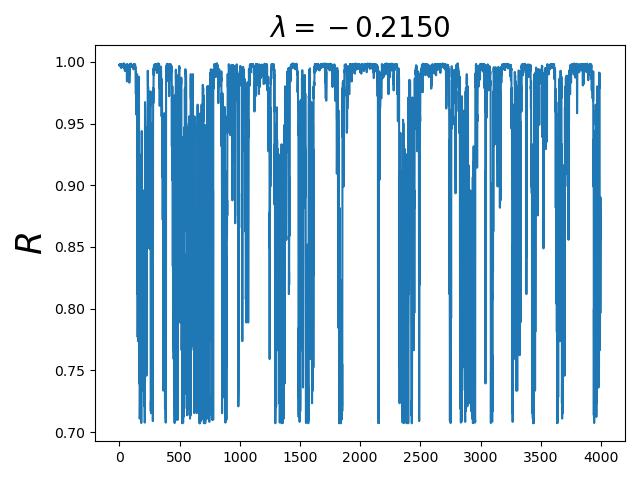}~~
      \includegraphics[width=0.3\textwidth]{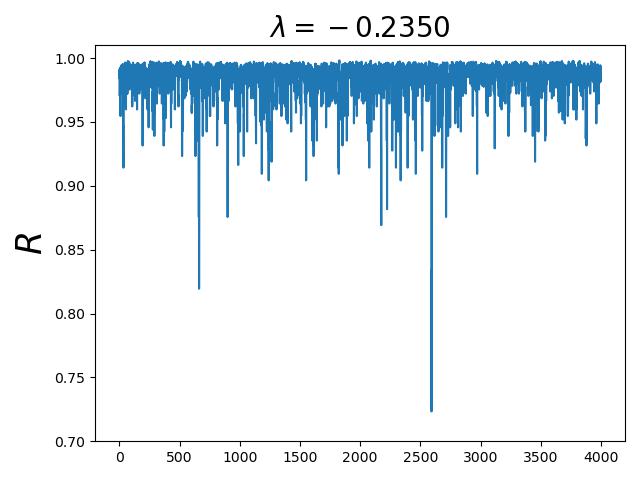}}
  \end{hbox}         
        \vskip-0.2cm
        \caption{The running histories of $R$ of sublattice B for $\lambda = -0.19$ (left), $\lambda = -0.215$ (middle), and $\lambda = -0.235$ (right).}
        \label{fig5}
\end{figure*}

The running histories of $R$ on sublattice $A$ for $\lambda = -0.19, -0.215$, and $-0.235$ on a 24 by 24 lattice are shown as the left, middle, and right
panels of fig.~\ref{fig4},
respectively. These results imply that the sublattice $A$ is ordered for
both the regions of $\lambda > \lambda_c$ and $\lambda < \lambda_c$.
In addition, the running history of $R$ at $-0.215$ (which is the expected
$\lambda_c$) shows a tunneling between $R=1$ and $R=1/\sqrt{2}$.
This is a feature of a first-order phase transition. Similarly,
The running histories of $R$ on sublattice $B$ demonstrated in fig.~\ref{fig5}
indicate that sublattice
$B$ is ordered for $\lambda < \lambda_c$ and disordered for $\lambda > \lambda_c$.
In particular, the middle panel of fig.~\ref{fig5} also shows a signal of
first-order phase transition. In other words, both sublattices $A$ and $B$ reveal
the fact that the phase transition is first order. Here we would like
to point out that with the traditional Monte Carlo approach,
the first-order signal appears only when the system sizes are sufficiently large
($L \ge 48, 64$). For the NN calculations, one obtains the correct conclusion
regarding the nature of the phase transition from the outcomes of $L=24$.

\section{Discussions and Conclusions}

In this study, we investigate the phase transition of the 2D $U(1)$ quantum link model
on the triangular lattice using the method of NN. The employed
NN has only one hidden layer of two neurons and is trained without any input from the considered model.
Because of this set up, the training process takes only 24 seconds on a server with two
opteron 6344 and 96G memory.

Despite its simplicity, with the magnitude $R$ of the output,
the constructed NN not only determines the critical $\lambda_c$
precisely, it also identifies the physics before and beyond the transition correctly.
These conclusions are obtained solely from the NN data without consulting any information about
the investigated model.

The conventional NNs used to study phase transitions are typically made up of several layers
and each layer has many neurons. Such a NN architecture leads to a lot of tunable parameters,
hence the corresponding training is very time-consuming. Intuitively, the idea of considering complicated
deep learning NN is that the studied systems have huge amounts of degrees of freedom,
therefore one needs to tune many NN parameters in order to capture the right physics.
Our investigation as well as that in Ref.~\cite{Hse22} suggest that one hidden layer
having two neurons is sufficient the uncover the phase structures of many nontrivial models.

To promote the adoption of NN techniques in real calculations, the NNs definitely should have some
advantage that makes NNs outperform the traditional approaches. When phase transition
is concerned, the conventional NNs considered in the literature require quite a lot of
computing efforts. It is then not so promising to use NN instead of the traditional
methods to perform the needed calculations. The unconventional NN strategy shown
here uses only several minutes to complete the whole calculation (from the training to the prediction).
As a result, it offers an efficient alternative to the traditional approaches when
phase transitions are studied.

Finally, similar to the unsupervised AE and GAN of Ref.~\cite{Hse22}, the supervised NN considered here
can directly applies to other models without performing any new training.

\section*{Acknowledgement}\vskip-0.3cm
Partial support from National Science and Technology Council (NSTC) of Taiwan
is acknowledged.

\end{document}